\documentstyle[12pt,epsfig]{article}

\begin{document}
\topmargin -1.4cm
\oddsidemargin -0.8cm

\evensidemargin -0.8cm 

\title{Non-Abelian bootstrap of primordial magnetism}

\vspace{1.5cm}

\author{Poul Olesen\footnote{email: 137olesen@gmail.com}\\
{\it  The Niels Bohr Institute}\\
{\it Blegdamsvej 17, Copenhagen \O, Denmark }}

\maketitle

\begin{abstract}
We point out that a primordial magnetic field can be generated in the
electroweak phase transition by a non-Abelian bootstrap, where the
field is generated by  currents of $W'$s, which in turn are 
extracted from the vacuum by the magnetic field. This magnetic field is produced
as a vortex condensate at the electroweak phase transition. It becomes
stringy as a consequence of the dynamical evolution due to magnetohydrodynamics.

\end{abstract}

\thispagestyle{empty}

\vskip0.5cm

There is much evidence for the existence of a primordial magnetic field.
There are several proposals for how these fields are generated, as reviewed
for example in the paper \cite{hector}. One possibility is genesis at the
electroweak phase transition, as first discussed by Vachaspati \cite{tand}.
In his case the magnetic field is generated from properties of the 
Higgs field. In this note we shall discuss another possible mechanism
for generation at the electroweak phase transition, namely a
non-Abelian bootstrap mechanism whereby a magnetic field is
generated from currents coming from charged $W'$s which in turn are
generated from the magnetic field. This self organized mechanism relies heavily
on properties that are generic for non-Abelian vector fields.
The proposed mechanism is therefore also of interest in principle,
since it may give direct observational information on non-Abelian field 
theory of vectors.

We start by considering a simple model with an SU(2) massive vector field,
\begin{equation}
{\cal L}=-\frac{1}{4}~F_{\mu\nu}^2-m_W(T)^2~W^\dagger_\mu W_\mu.~~W_\mu=
\frac{1}{\sqrt{2}}~(A_\mu^1+iA_\mu^2),
\label{model}
\end{equation}
where
\begin{equation}
F_{\mu\nu}=\partial_\mu   A_\nu-\partial_\nu A_\mu +ig[A_\mu,A_\nu],~~A_\mu=
A_\mu^a ~
\sigma_a/2. 
\end{equation}
This is the same model as considered some time ago by Ambj\o rn and the
author in \cite{amb}, except that the mass depends
on the temperature $T$. It is assumed that there is a phase transition,
\begin{equation}
m_W(T)=0~~{\rm for}~~T>T_c,~~m_W(T)\neq 0~~{\rm otherwise.} 
\end{equation}
The magnetic field
\begin{equation}
f_{\mu\nu}=\partial_\mu A_\nu^3 -\partial_\nu A_\mu^3
\label{f}
\end{equation}
is given by \cite{amb}
\begin{equation}
ef_{12}=m_W(T)^2+2e^2~|W|^2,
\label{field}
\end{equation}
where we used the ansatz \cite{amb} $W=W_1,~W_2=iW_1\equiv iW, W_3=W_0=0, 
W=W(x_1,x_2)$. The 
result (\ref{field}) arizes from minimizing  the energy written in the form
\begin{equation}
{\rm Energy~density}=
|(D_1+iD_2)W|^2+\frac{1}{2}\left(f_{12}-\frac{m_W(T)^2}{e}-2e|W|^2\right)^2+
\frac{m_W(T)^2}{e}f_{12}-\frac{m_W(T)^4}{2e^2 }.
\end{equation}
In this equation
\begin{equation}
D_a=\partial_a-ieA_a^3.
\end{equation}

We carry out the  minimization by requiring that the two positive quadratic 
terms vanish 
like in the Bogomol'nyi limit.
We immediately obtain Eq. (\ref{field}). The equation of 
motion for $W=|W|\exp(i\chi)$ can be obtained from the vanishing of 
the first term,
\begin{equation}
(D_1+iD_2)W=0,
\end{equation}
from which we get by use also of Eq. (\ref{field})
\begin{equation}
-(\partial_1^2+\partial_2^2)\ln |W|=m_W(T)^2+2e^2|W|^2-\epsilon_{ij}\partial_i
\partial_j\chi,
\label{eqofmotion}
\end{equation}
where $\chi$ is the phase of $W$. The relative plus sign between the
two terms on the right hand side
of Eq. (\ref{field}) reflects the antiscreening\footnote{This is the 
anti-Lenz' law according to which the magnetic field will be enhanced
by the current. The necessary energy is produced by the $W-$condensate.}  
of this solution. Because of 
this sign  there is no single vortex solution. Instead
Eq. (\ref{eqofmotion}) has periodic  solutions, corresponding to a 
lattice of
vortices. As usual in each periodicity domain the term 
$\epsilon_{ij}\partial_i\partial_j\chi$ gives a 
delta function corresponding to the delta function coming from the zero of 
$|W|$ on the left hand side of (\ref{eqofmotion}).

The  result (\ref{field}) is now considered as a self organized
solution of the equations of motion which shows the possibility of
creating a magnetic field from the non-Abelian dynamics. The energy is taken 
from the expansion energy of the universe. 

The order of magnitude of the field below the temperature $T_c$ is 
\begin{equation}
f_{12}\sim m_W^2/e\sim 10^{24}~G,
\end{equation}
which is a large field, of the same order of magnitude as the one
found by Vachaspati \cite{tand}. Each flux tube has a dimension of order
$1/m_W$. For $T>T_c$ the solution of the equation of motion
(\ref{eqofmotion}) can be found explicitly in terms Weierstrass'
p-function, and it can be shown that it corresponds to a zero
energy solution \cite{po}. By a nonperturbative gauge transformation
one can transform this solution to the perturbative ground state
$A_\mu^a=0$. 

The solution of Eq. (\ref{eqofmotion}) is a bootstrap type of
solution, because the magnetic field is inherent in the vacuum and
is extracted from ``emptiness'' by the appearance of the mass in the phase  
transition and is kept alive by currents from the $W'$s. In other words, the 
magnetic field is generated by $W-$currents, 
\begin{equation}
\partial_1f_{12}=2e\partial_1|W|^2=-j_2~~{\rm and}~~\partial_2f_{12}=
2e\partial_2|W|^2=j_1,
\end{equation}
and these $W'$s in turn are 
generated by the magnetic field,
because of the non-Abelian  instability discussed a long time ago \cite{nkn},
according to which the magnetic field exceeding the magnitude $m_W^2/e$ is
unstable unless stabilized by $W'$s from the vacuum. Thus the magnetic field
and the vector bosons are interwoven in the structure of the solution of
Eq. (\ref{eqofmotion}) and only exist because of one another.

The energy density is given by
\begin{equation}
{\cal E}=\frac{m_W(T)^2}{e}~f_{12}-\frac{1}{2}~\frac{m_W(T)^4}{e^2}
\end{equation}
We see that this energy is smaller than the no condensate energy $f_{12}^2/2$ 
due to the negative contribution from the $W$ condensate.

Considering the vortices as strings we can compute the string tension by 
integrating the energy density over a single quadratic\footnote{For simplicity
we consider a quadratic domain. Energy may be minimized by other types of 
domains. The constant  $c$ is close to $2\pi$
if $m_W^2>>2e²|W|^2$.} domain with 
area $c/m_w^2$, where $c$ is a numerical constant. The result is
\begin{equation}
{\rm string~tension}\equiv\sigma (T)=(2\pi-c/2)~\frac{m_W(T)^2}{e^2},
\label{tension}
\end{equation}
where we used the quantization of the flux
\begin{equation}
{\rm Flux}=\int_{\rm domain}f_{12}d^2 x=2\pi/e.
\end{equation}
Thus we see that the string tension\footnote{In a certain
non-perturbatively defined gauge the strings still exist \cite{po} with
zero string tension above
the critical temperature. This vacuum
string configuration is, however, degenerate with the perturbative vacuum.}
 vanishes above the critical temperature,
where the field contents of the solution becomes non-perturbative pure 
gauge fields, as discussed in \cite{po}.

As the universe expands the strings develop according to the 
magnetohydrodynamic (MHD) field equations. A long time ago we showed 
\cite{mhdpo} that in the limit of infinite conductivity (``ideal'' MHD) these
equations are satisfied by Nambu-Goto strings. Later this was
discussed including dissipative effects by Schubring \cite{schubing}.
Also,  numerically the turbulent plasma governed by the MHD equations 
has been found to be extremely intermittent with
the vorticity concentrated in thin vortex types with the magnetic field 
concentrated also in thin vortex types \cite{aake}. Therefore the stringy 
initial behavior exhibited
by the vortex condensate discussed above fits well with the
subsequent MHD governed develpment of the universe. In the string picture the 
magnetic field is given by \cite{mhdpo},\cite{schubing}
\begin{equation}
B_i({\bf x},t)=\sum_{\rm strings}b\int d\sigma \frac{\partial z_i(\sigma,y)}
{\partial\sigma}~
\delta^3 (x-z(\sigma,t)),
\end{equation}
where $f_{12}=B_3$ etc. and $b$ is the magnetic flux. In this equation there 
should be a sum over all the
strings in  the vortex lattice. The string coordinates
satisfy 
\begin{equation}
\frac{\partial z_i}{\partial t}\frac{\partial z_i}{\partial \sigma}=0,~~
{\rm and}~~\frac{\partial^2z_i}{\partial \sigma^2}=\frac{1}{v_0^2}
\frac{\partial^2z_i}{\partial t^2}. 
\end{equation}
Here $v_0$ is a maximum transverse velocity.

The string tension from the gauge theory is temperature dependent and vanishes 
above the critical 
temperature. This phenomenon was found in string theory for the Nambu-Goto 
string long time ago by Pisarski and Alvarez \cite{pisarski}, where the 
critical temperature is the deconfinement (Hagedorn) temperature. 
Their result would be obtained from Eq. (\ref{tension}) if 
\begin{equation}
m_W(T)^2\propto \sqrt{1-(T/T_c)^2}
\end{equation}
In our case, the exitence of the critical 
temperature indicates that the vortex/string picture breaks down above this
temperature. Of course, even in the field theory case this temperature would 
also correspond to  deconfinement
if monopoles exist. They would be confined below the critical
temperature, and released above the critical temperature because of zero string
tension. 

So far we have consideed the simple model (\ref{model}) with a temperature 
dependent mass. We shall now consider the standard electroweak theory with a
Higgs field $\phi$, where the magnetic field turns out to be given by \cite{ew}
\begin{equation}
f_{12}=\frac{g~\phi_0(T)^2}{2\sin \theta}+2g\sin\theta |W|^2
\end{equation}
in the Bogomol'nyi limit where the Higgs mass equals the $Z$ mass. For the 
realistic mass case a much more complicated perturbative treatment is
necessary. For
simplicity we shall therefore stick to the Bogomol'nyi limit. The equations 
of motion 
are \cite{ew} 
\begin{equation}
-(\partial_1^2+\partial_2^2)\ln |W|=\frac{g^2}{2}\phi^2+2g^2|W|^2-
\epsilon_{ij}\partial_i\partial_j\chi,
\end{equation}
which is analogous to Eq. (\ref{eqofmotion}), and
\begin{equation}
(\partial_1^2+\partial_2^2)\ln\phi=\frac{g^2}{4\cos^2\theta}~(\phi^2-\phi_0(T)^2)
+g^2|W|^2.
\end{equation}
It has been proven mathematically that these coupled equations have
periodic solutions \cite{math}-\cite{math5}. Here $\phi_0(T)$ vanishes 
above the critical temperature and the solution then becomes degenerate 
with the perturbative vacuum \cite{po}.

The string tension can again be obtained by integrating the energy density
over one domain in the plane. The result is the same as in
Eq. (\ref{tension}). Thus the previous discussion can be repeated for
the electroweak theory, at least in the Bogomol'nyi limit. Again the
string tension vanishes above the critical temperature.

We end this discussion with some remarks on the chiral anomaly effect
on the evolution of the primordial magnetic field 
\cite{chiral1}-\cite{chiral6}. The inclusion of this effect will
modify the MHD equations by adding an  effective electric current.
Also, hyperfields are relevant above the electroweak phase transition,
and there may be magnetic helicity above and below this transition. In ideal
MHD helicity is conserved, but this is not valid when the Ohmic resistance
is included, and the helicity will ultimately decay. It is clear that our 
solution is not born with helicity, since for this solution 
${\bf AB}=0$, but due to fluctuations from the full MHD equations there 
will always be some helicity \cite{it}.

It is always a problem for primordial magnetic fields generated 
from particle physics that the initial scale is small. Even though
the expansion of the universe increases this scale in general this is
not enough for the generation of realistic scales. Therefore the phenomenon
of inverse cascading, i.e. the drift of energy towards larger scales,
 is important \footnote{A more appropriate term is perhaps drift
towards the infrared, i.e. small $k$.}. Often this phenomenon is linked with 
(conserved) helicity \cite{h}.
However, with vanishing helicity 
there is still an inverse cascade in freely decaying MHD, moving energy from 
smaller
to larger scales, as discussed recently \cite{ic1}-\cite{ic3}. Thus,
helicity is {\it not} a necessary condition for an inverse cascade to occur. 
More explicitly it  was  found numerically
by Zrake \cite{ic1} that the energy scales in a self-similar manner, which
was shown by the author \cite{ic3} to be an {\it exact}
consequence of the standard MHD equations for freely decaying
turbulence when  dissipation is included.
The energy density should satisfy
\begin{equation}
{\cal E}(k,t)=\sqrt{\frac{t_0}{t}}~{\cal E}
\left(k\sqrt{\frac{t}{t_0}},t_0\right)
\label{scale}
\end{equation}
According to this formula (which is one of the few known exact results in
HD and MHD) energy is moved from smaller to larger 
scales as time passes. This 
is obviously important in order to increase the scale of the magnetic 
field on the top of the expansion of the universe. These results have 
to be modified if chiral MHD turbulence 
is taken into account, as discussed recently in \cite{chiral6}.

For completeness we display the magnetic energy for an expanding flat universe
with  the metric
\begin{equation}
d\tau^2=dt^2-a(t)^2d{\bf x}^2=a(\tilde{t})^2\left(d\tilde{t}^2-
d{\bf x}^2\right).
\end{equation}
Here $t$ is the Hubble time and
\begin{equation}
\tilde{t}=\int dt/a(t)
\end{equation}
is the conformal time corresponding to the expansion parameter $a(t)$. Eq.
(\ref{scale}) is then replaced by
\begin{equation}
{\cal E }_B(k,\tilde{t})=\left(\frac{a(\tilde{t_0})}{a(\tilde{t})}\right)^4~
\sqrt{\frac{\tilde{t_0}}{\tilde{t}}}~{\cal E}_B
\left(k\sqrt{\frac{\tilde{t}}{\tilde{t_0}}},\tilde{t_0}\right).
\end{equation}
Again we see a drift towards large distances as the univese expands.

\vskip0.5cm

In conclusion we have shown that a primordial magnetic field can be
generated in the electroweak phase transition by a non-Abelian vector
bootstrap. The resulting field consists of a set of antiscreening
vortices which,
because they have to follow the MHD equations, develop in a stringy manner. 
Due to the inverse cascade the field may survive at large distances so as to
have a realistic scale at
the present time. If it turns out that the relevant field is generated 
as a hyperfield above the electroweak phase
transition it may still receive additional contributions from the
bootstrap mechanism discussed here when the universe passes the 
electroweak phase transition and the hyperfield turns into a magnetic
field.

\end{document}